\documentclass[prb,aps,preprint]{revtex4-1}
\usepackage{graphicx}
\usepackage{bm}
\addtocounter{section}{-1}
\begin{document}

\begin{center}

\vspace{2cm}

{\large {\bf {
Possible magnetic states in buckybowl molecules
} } }

\vspace{1cm}

{\rm Kikuo Harigaya\\$_{\rm E-mail address: 
k.harigaya@aist.go.jp; URL: 
http://staff.aist.go.jp/k.harigaya/}$}

\vspace{1cm}

{\sl Nanosystem Research Institute, AIST, 
Tsukuba 305-8568, Japan}\\

\vspace{1cm}

(Received~~~~~~~~~~~~~~~~~~~~~~~~~~~~~~~~~~~)
\end{center}

\vspace{1cm}

\noindent
{\bf Abstract}\\
Possible magnetic properties are studied in the buckybowl
molecules: the sumanene and a part of C$_{60}$.  The Hubbard
model is applied to the systems.  We find that the molecular 
structure determines the magnetism in the sumanene.  
On the other hand, the edge state is found along the zigzag 
edge of a part of C$_{60}$.  Therefore, the novel property, 
transition from molecular magnetism to the magnetism like
in nanographene, is found.

\vspace{1cm}


\pagebreak

\section{Introduction}

Nanographene materials [1,2] have been studied intensively.
The recent Nobel prize paper [3] on graphene has promoted
experimental and theoretical investigation on nanocarbon
materials, too.  Unique magnetic properties along the zigzag
edge of the graphene nanoribbon [1] have been predicted.
On the other hand, defects on the fullerene C$_{60}$ [4] 
may give possible unique electronic states.  Further, we 
have studied magnetic properties of stacked nanographene sheets
by the cluster calculations with the Hubbard model [5-7].

The buckybowl molecule ''sumanene" [8,9] has been synthesized
experimentally.  The molecule has a curved shape, and is a
part of the structure of C$_{60}$.  The molecule has $\pi$-electrons
on the curved surface.  The electronic properties have been
calculated [10], and their uniqueness has been revealed.
Therefore, it is also interesting to study possible magnetic
properties.  In addition, we consider a part of C$_{60}$,
where three pentagons and six hexagons are added around
the outer part of the sumanene.  A zigzag line appears, 
so magnetic properties can be compared with those of
graphene nanoribbons [1].

In this paper, possible magnetic properties are studied in 
the buckybowl sumanene and a part of C$_{60}$ with the 
zigzag edge like in nanographene.  The Hubbard model [5-7] 
is applied to $\pi$-electrons on the curved molecules.  
We will discuss that the molecular structure determines 
the magnetism in the sumanene, while the edge state gives
magnetism along the zigzag edge of a part of C$_{60}$.  
Therefore, the novel property, transition from molecular 
magnetism to the magnetism in nanographene, appears.

This paper is organized as follows.  In section 2, we explain
the model.  In section 3, we report magnetism of the sumanene.
In section 4, we discuss about a part of C$_{60}$.
The paper is closed with a short summary in section 5.

\section{Model and method}

We treat half-filled $\pi$-electron systems on buckybowl
molecules by the Hubbard model.  This is a well known
model for description of localized magnetic properties
of itinerant electrons.  The model is as follows:
\begin{equation}
H = - t \sum_{\langle i,j \rangle,\sigma}
( c_{i,\sigma}^\dagger c_{j,\sigma} + {\rm h.c.} )
+ U \sum_{i} 
(c_{i,\uparrow}^\dagger c_{i,\uparrow} - \frac{n_{\rm el}}{2})
(c_{i,\downarrow}^\dagger c_{i,\downarrow} 
- \frac{n_{\rm el}}{2}),
\end{equation}
where $c_{i,\sigma}$ annihilates a $\pi$-electron of 
spin $\sigma$ at the $i$th site; $t$ ($> 0$) is the hopping 
integral between the nearest neighbor $i$th and $j$th sites; 
the sum with $\langle i,j \rangle$ is taken for 
all the pairs of the nearest neighbor sites;  $n_{\rm el}$ is 
the average electron density of the system.  Here, $n_{\rm el}=1$
for the half filled electrons.  The term $U$ is the strength
of repulsive force between up and down spins at the $i$th
site.  When $U$ is larger than $t$, the electrons favor
to localize, and magnetic states are realized.
When $t$ is larger than $U$, the magnetic character
is suppressed due to the strong itinerant property.
We adopt the Hartree-Fock approximation to this model [5-7]:
\begin{equation}
c_{i,\uparrow}^{\dagger}c_{i,\uparrow} 
 c_{i,\downarrow}^{\dagger}c_{i,\downarrow} 
\Rightarrow \langle c_{i,\uparrow}^{\dagger}c_{i,\uparrow}\rangle
c_{i,\downarrow}^{\dagger}c_{i,\downarrow}
+c_{i,\uparrow}^{\dagger}c_{i,\uparrow}
\langle c_{i,\downarrow}^{\dagger}c_{i,\downarrow}\rangle
-\langle c_{i,\uparrow}^{\dagger}c_{i,\uparrow}\rangle
\langle c_{i,\downarrow}^{\dagger}c_{i,\downarrow}\rangle.
\end{equation}
The magnetic moment at the $i$th lattice site is
calculated as:
\begin{equation}
m_i = \frac{1}{2} 
(\langle c_{i,\uparrow}^\dagger c_{i,\uparrow} \rangle
- \langle c_{i,\downarrow}^\dagger c_{i,\downarrow} \rangle).
\end{equation}

\section{Magnetic states of sumanene}

The sumanene molecule has a curved shape, where a hexagon
is located at the center of the molecule.  Figure 1 shows
the molecular structure.  The pentagons and hexagons are
alternated among the central hexagon.  The molecule has 
the C$_3$ symmetry.  The central hexagon has bond alternation
of the single and double bonds.  Three surrounding hexagons
have two double bonds.  There are three nonequivalent edge
atoms due to the symmetry.  They are named as A, B, and C.
It is assumed that $\pi$ electron is active at the site A.
In other words, the site A is terminated with one hydrogen
atom, not with two H atoms.  This assumption is necessary
for the appearance of the magnetism.

The Hartree-Fock equation is solved with a certain initial 
condition for the electron number 21.  This is the same as 
the carbon atom number.  The magnetic solution is actually 
obtained. The magnetic moments at the sites A, B, and C
are plotted against the Coulomb interaction strength $U$
in Fig. 2.  The interaction is changed within $0 < U < 2.5t$.
The magnetic moment appears at the site A [Fig. 2 (a)].  
Its magnitude increases from about 0.2 of $U=0$, as the 
strength $U$ becomes larger.  The magnitude of the magnetic 
moments at the sites B and C [Figs. 2 (b) and (c)] is 
one-order smaller than that of the site A.  The existence
of magnetic moments at the sites B and C will not be
observed.  The appearance of the moment at the site A
is related with the fact that the site is the outmost
site of the pentagon ring of the molecule.  On the other
hand, the bond between the sites B and C is a double
bond, so the singlet spin state is favorable and
localized magnetic moments will disappear.
This type of magnetic state reflects the alternation
pattern of the single and double bonds, so it
is one of molecular magnetic states.

Such molecular magnetism is closely related with the
molecular magnetism in the trianglene [11,12].
The nonbonding molecular orbitals give rise to
the magnetic nature of the trianglene, too.
Possibly, electron-phonon couplings may give
rise to distortions to molecular structures,
namely the Jahn-Teller distortions.  However,
in the stage of this simple tight binding
model, we do not have detailed knowledge of
electron-phonon (molecular vibrations) 
couplings.  Thus, we have limited the work
to the magnetic solution of the Hubbard model.
Inclusion of the molecular distortion can
be done possibly by using the Su-Schrieffer-Heeger
type interaction performed for the works of 
carbon nanotubes [13] and graphene frakes [14].

\section{Magnetic states in a part of C$_{60}$}

In this section, we consider magnetic states in a part 
of C$_{60}$, where three pentagons and six hexagons are 
added around the outer part of the sumanene [Fig. 1].  The 
new structure is shown in Fig. 3.  The structural relation with 
the sumanene would be obvious. A zigzag line appears
along the outmost part of the molecular structure, 
so magnetic properties can be compared with those of
graphene nanoribbons [1].  This is another interest
because bulk magnetic states have been predicted
in graphene nanoribbons.  Along the zigzag line of Fig. 3,
there are four nonequivalent edge sites.  They are
named sites A to D.  The site A is at the outmost site
of the pentagon, and sites B, C, and D belongs to hexagons.

The calculated result of the magnetic state due to
the Hartree-Fock approximation is summarized in Fig. 4.
The magnetic moment at the site A (a), site B (b),
site C (c), and site D (d) is displayed as a function 
of the Coulomb interaction strength $U$.  We find the switch-on
of the positive magnetic moments at the sites A and C, and
the presence of the negative moments at the sites B and D.  The 
absolute values of the moments increase as the strength $U$
becomes larger.  The moments of the sites A and C seem 
to be finite at $U=0$, because these sites locate at the
outmost part of the molecule.  On the other hand,
the moments of the sites B and D are near to zero
due to the three bonded nature of these sites.
Such the alternation of the positive and negative
magnetic moment along the zigzag edge of a part
of C$_{60}$ would be by no means the realization 
of the edge state magnetism, which has been discussed
in graphene nanoribbons [1].  The electrons of the
magnetic moments occupy wavefunctions which have 
large amplitudes at and near the zigzag atoms
in these systems.

At the first glance, the above qualitative behavior of the 
magnetic property seems similar with that of the nanographene.
However, there is a quantitative difference.  In nanographene
ribbon with zigzag edges [1], the magnetic moment at the edge 
atoms (types of the sites A and C of Fig. 3) is larger than that
at the neighboring sites (types of the sites B and D).  So, 
ferrimagnetism occurs.  In the present system, the absolute value 
of the magnetic moment of the sites A to D is of the similar
magnitude.  This is the quantitative difference from
the graphene nanoribbons.  In density functional calculations [15],
magnetic moments at edge atoms are displayed only.
It is possible that the similar alternation of the
up and down spins has been obtained in the {\sl ab initio} study [15].
The recent molecular dynamics simulations [16] correspond
to this calculation, and stabilities of the molecules and
structural transitions have been discussed.

Comparing magnetic properties between the sumanene and a
part of C$_{60}$, the magnetic moment is suppressed at
the sites B and C of the hexagons of the sumanene in Fig. 1,
while magnetic moments are favored at the sites B, C, and D
of the hexagons in Fig. 3.  The magnetism of the sumanene
is of the molecular nature [11,12], while that of the part of C$_{60}$
is like of the bulk magnetism of nanographene [1].
Therefore, by changing the system size, transition from
molecular magnetism to magnetism like nanoribbons with
zigzag edges has been found.

It is known that the Hartree-Fock approximation tends 
to overestimate the possibility of broken symmetry magnetic 
solutions.  Fluctuations from the mean field approximation
could change the stabilities of solutions.  More precise
treatments may be done by quantum Monte Carlo or exact
diagonalization method, as has been used for the extended
Hubbard model on the C$_{20}$ molecule recently [17],
for example.  This would be one of the extensions to the future studies.
This paper has provided an indication of the trend of change
of magnetism when going from small molecules like the sumanene 
to larger ones like graphene nanoribbons.

\section{Summary}

In summary, possible magnetic orders have been studied 
in the buckybowl molecules: the sumanene and a part of C$_{60}$.  
The Hubbard model has been applied to the systems.  It has
been found that the molecular structure determines the magnetism 
in the sumanene.  On the other hand, the edge state is realized 
along the zigzag edge of a part of C$_{60}$.  Therefore, the 
novel property, transition from molecular magnetism to the 
magnetism in nanographene, has been found.

\noindent
{\bf Acknowledgement}\\
Useful discussion with A. Yamashiro (Kyoto University), 
S. Abe (AIST), and Y. Shimoi (AIST) is acknowledged.

\pagebreak
\begin{flushleft}
{\bf References}
\end{flushleft}

\noindent
$[1]$ M. Fujita, K. Wakabayashi, K. Nakada, 
and K. Kusakabe, J. Phys. Soc. Jpn. {\bf 65}, 1920 (1996).\\
$[2]$ K. Wakabayashi, M. Fujita, H. Ajiki, and M. Sigrist,
Phys. Rev. B {\bf 59}, 8271 (1999).\\
$[3]$ K. S. Novoselov, A. K. Geim, S. V. Morozov, D. Jiang, 
Y. Zhang, S. V. Dubonos, I. V. Grigorieva, and A. A. Firsov,
Science {\bf 306}, 666 (2004).\\
$[4]$ K. Harigaya, J. Phys.: Condens. Matter {\bf 4},
6769 (1992).\\
$[5]$ K. Harigaya, J. Phys.: Condens. Matter {\bf 13},
1295 (2001).\\
$[6]$ K. Harigaya, Chem. Phys. Lett. {\bf 340},
123  (2001).\\
$[7]$ K. Harigaya and T. Enoki, Chem. Phys. Lett. {\bf 351},
128 (2002).\\
$[8]$ H. Sakurai, T. Daiko, and T. Hirao, Science {\bf 301},
1878 (2003).\\
$[9]$ S. Higashibayashi and H. Sakurai, J. Am. Chem. Soc.
{\bf 130}, 8592 (2008).\\
$[10]$ U. D. Priyakumar and G. N. Sastry, J. Phys. Chem. A {\bf 105},
4488 (2001).\\
$[11]$ J. Inoue {\sl et al.}, J. Am. Chem. Soc. {\bf 123},
12701 (2001).\\
$[12]$ K. Fukui {\sl et al.}, Synth. Metals {\bf 121},
1824 (2001).\\
$[13]$ O. \"{O}zsoy, J. Optoelectron. Adv. Mater. {\bf 9}, 2283 (2007).\\
$[14]$ O. \"{O}zsoy and K. Harigaya, J. Comp. Theor. Nanoscience,
{\bf 8}, 31 (2011).\\
$[15]$ Y. H. Kim, J. Choi, K. J. Chang, and D. Tom\'{a}nek,
Phys. Rev. B {\rm 68}, 125420 (2003).\\
$[16]$ A. Hussien, A. V. Yakubovich, and A. V. Solov'yov,
AIP Conf. Proc. {\bf 1197}, 152 (2009).\\
$[17]$ F. Lin, E. S. S$\o$rensen, C. Kallin, and A. J. Berlinsky,
J. Phys. Condens. Matter {\bf 19}, 456206 (2007).\\

\pagebreak

\begin{flushleft}
{\bf Figure Captions}
\end{flushleft}

\mbox{}

\begin{figure}[t]
\centerline{
\includegraphics[height=7cm]{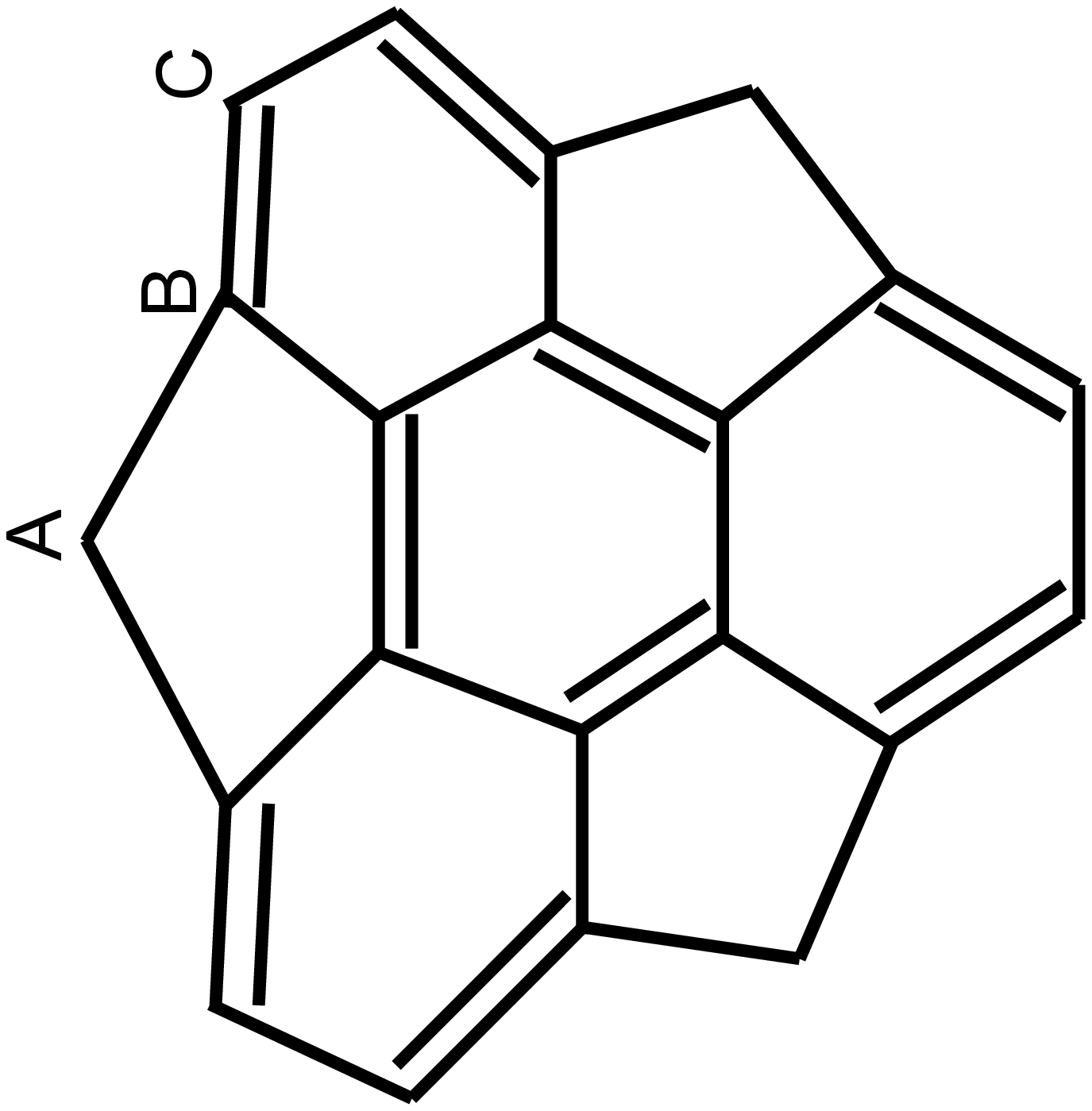}
}
\caption{
The sumanene molecule.  The symbols, A, B, and C,
label symmetrically unequivalent edge atoms.
}
\label{fig:Fig1}
\end{figure}

\mbox{}

\begin{figure}[t]
\centerline{
\includegraphics[height=5cm]{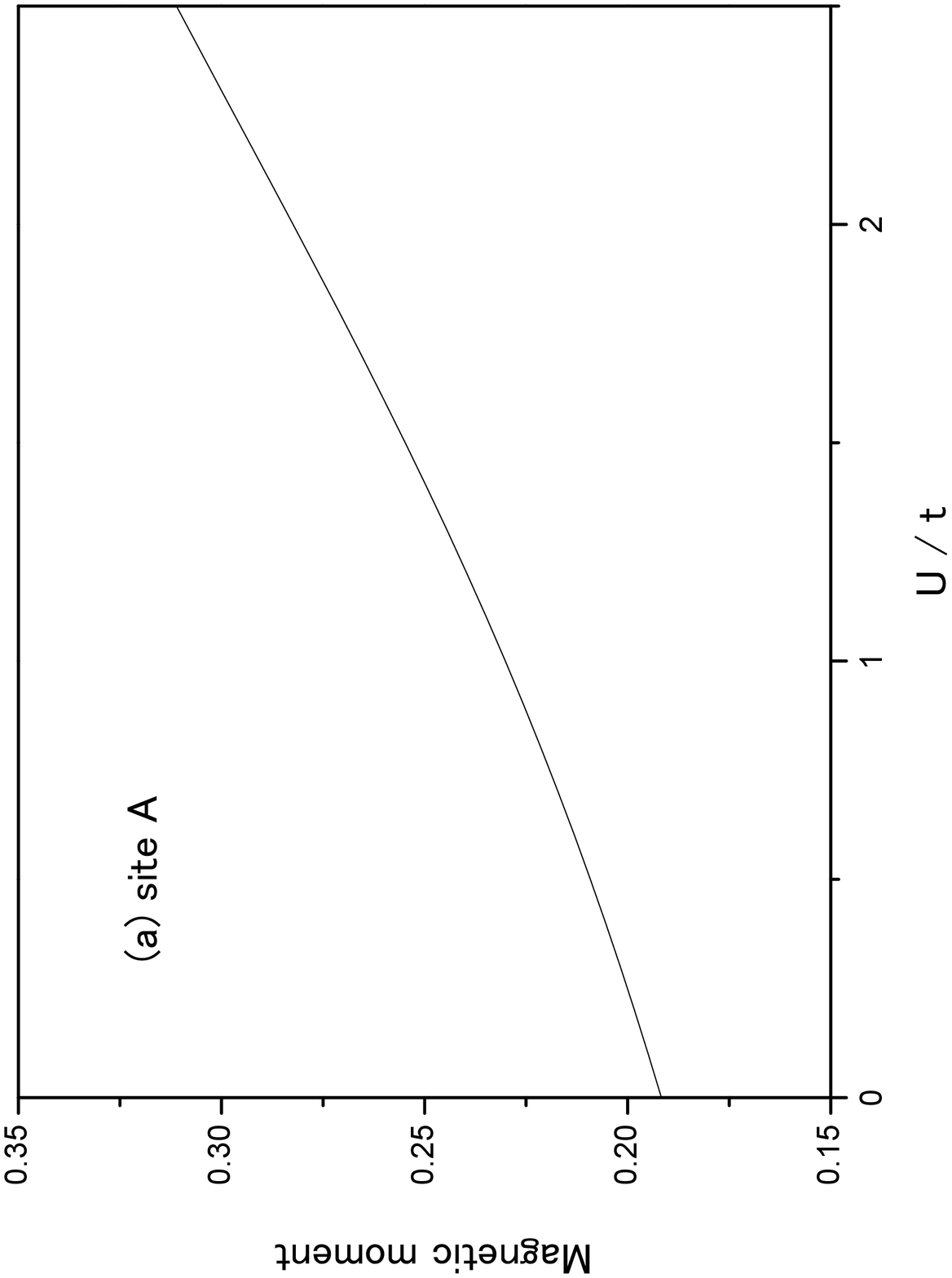}
}
\centerline{
\includegraphics[height=5cm]{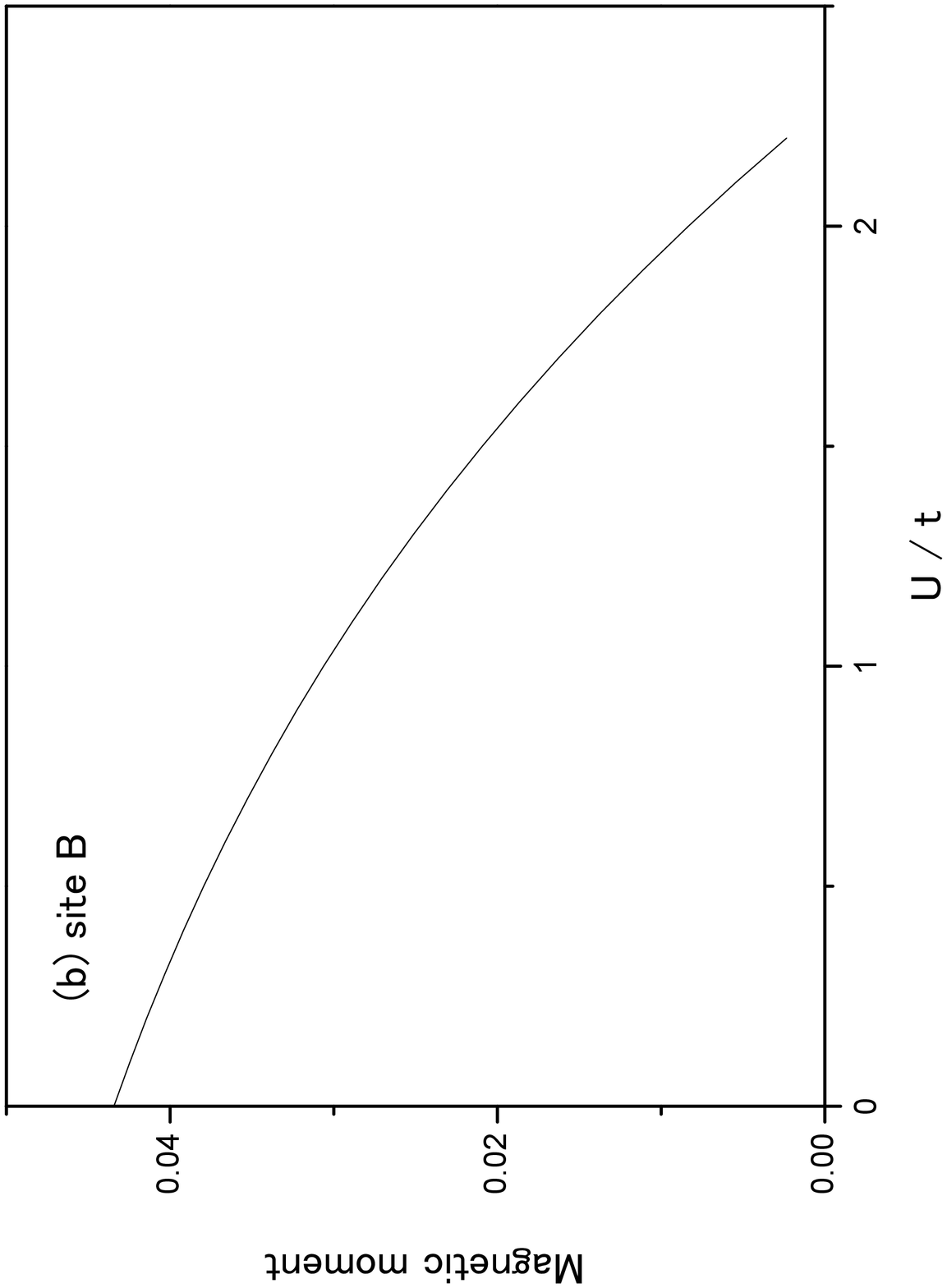}
}
\centerline{
\includegraphics[height=5cm]{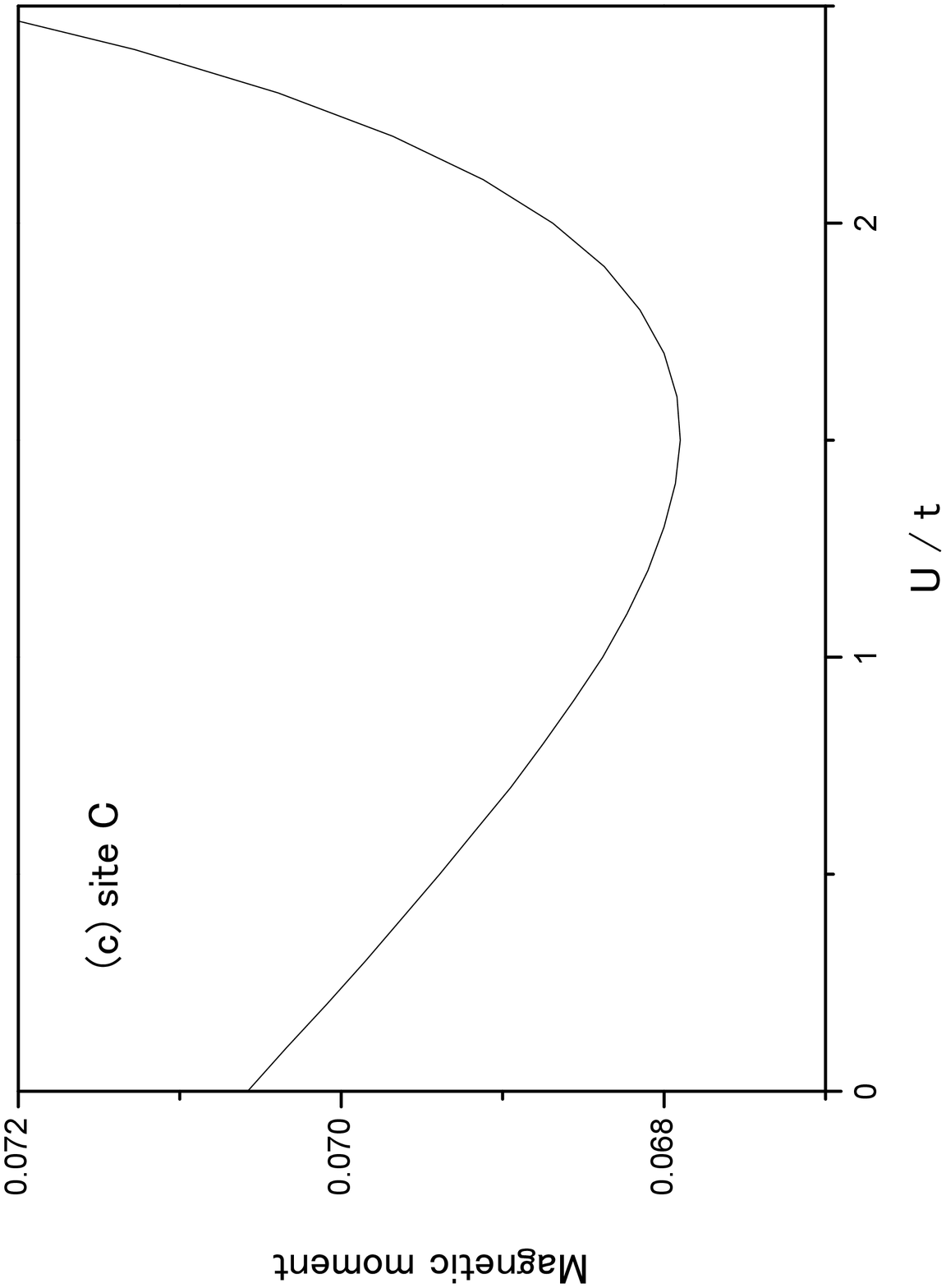}
}
\caption{
The magnetic moment at the site A (a), site B (b),
and site C (c), as a function of the Coulomb interaction 
strength $U$ of the sumanene molecule.
}
\label{fig:Fig2}
\end{figure}

\mbox{}

\begin{figure}[t]
\centerline{
\includegraphics[height=7cm]{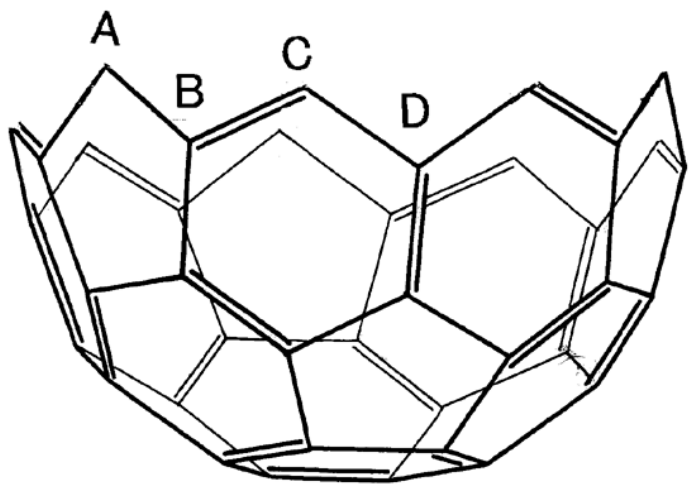}
}
\caption{
A part of C$_{60}$ with a zigzag line.  The symbols, 
A, B, C, and D, label symmetrically unequivalent edge atoms.
}
\label{fig:Fig3}
\end{figure}

\mbox{}

\begin{figure}[t]
\centerline{
\includegraphics[height=5cm]{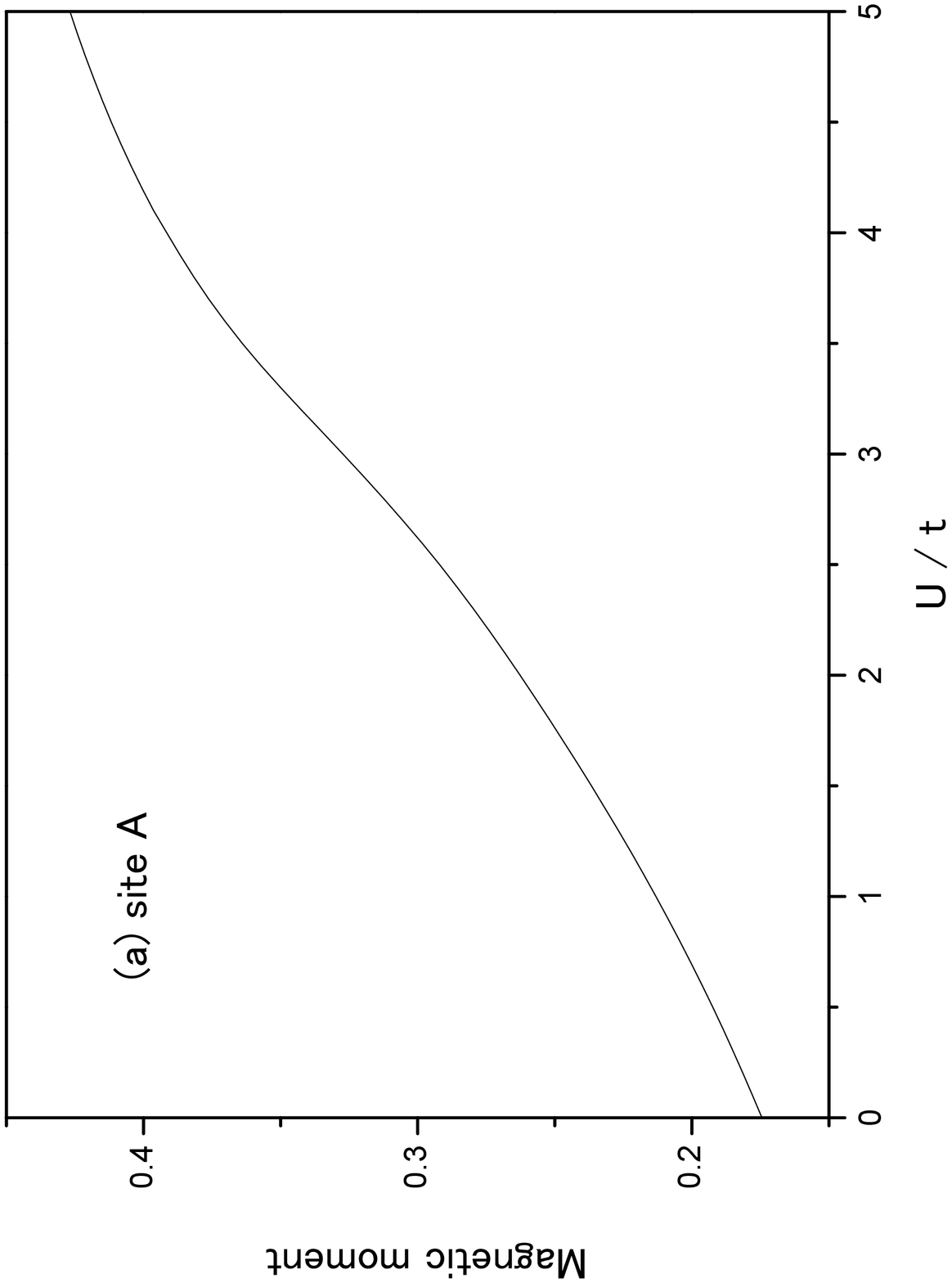}
}
\centerline{
\includegraphics[height=5cm]{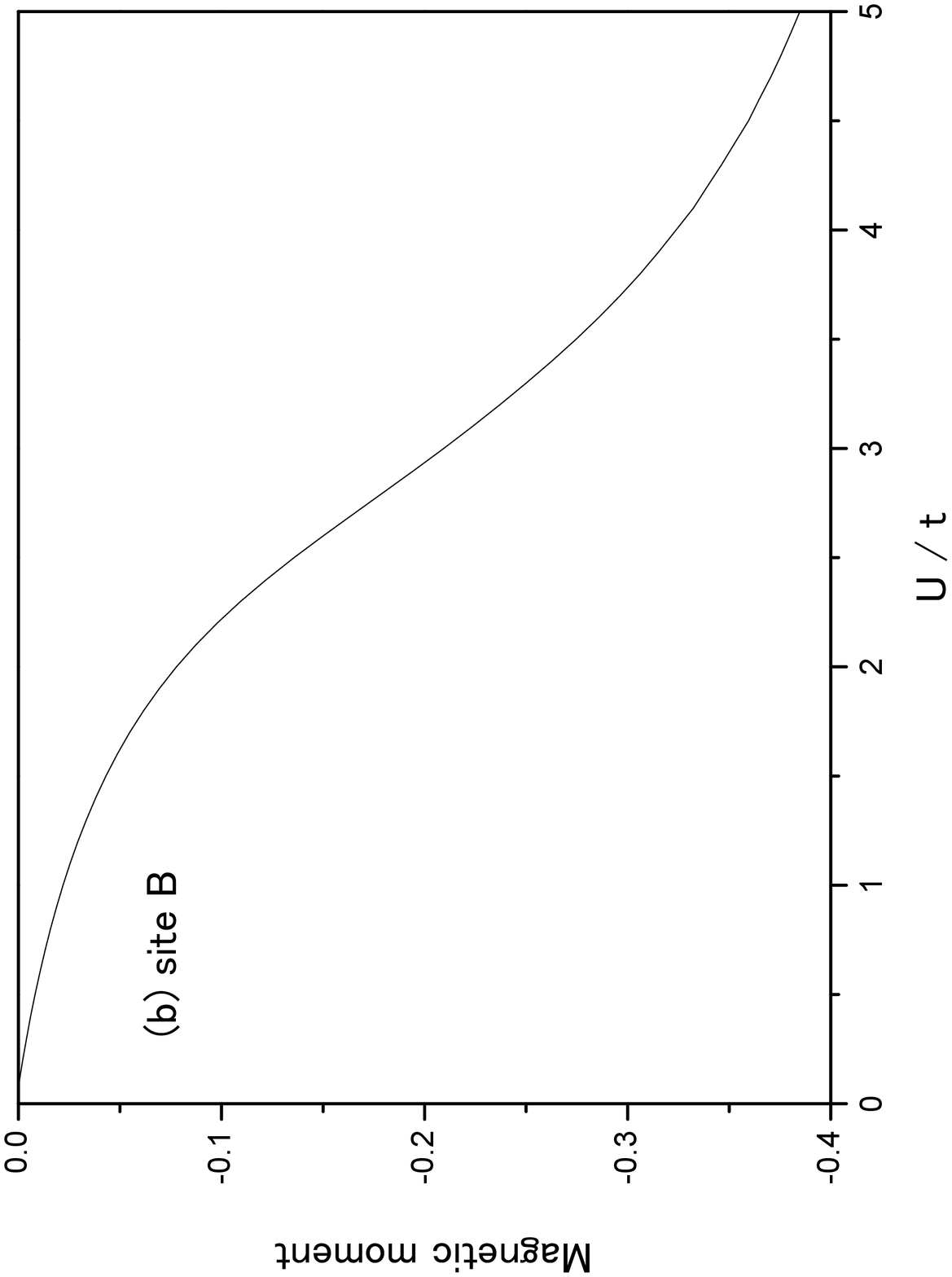}
}
\centerline{
\includegraphics[height=5cm]{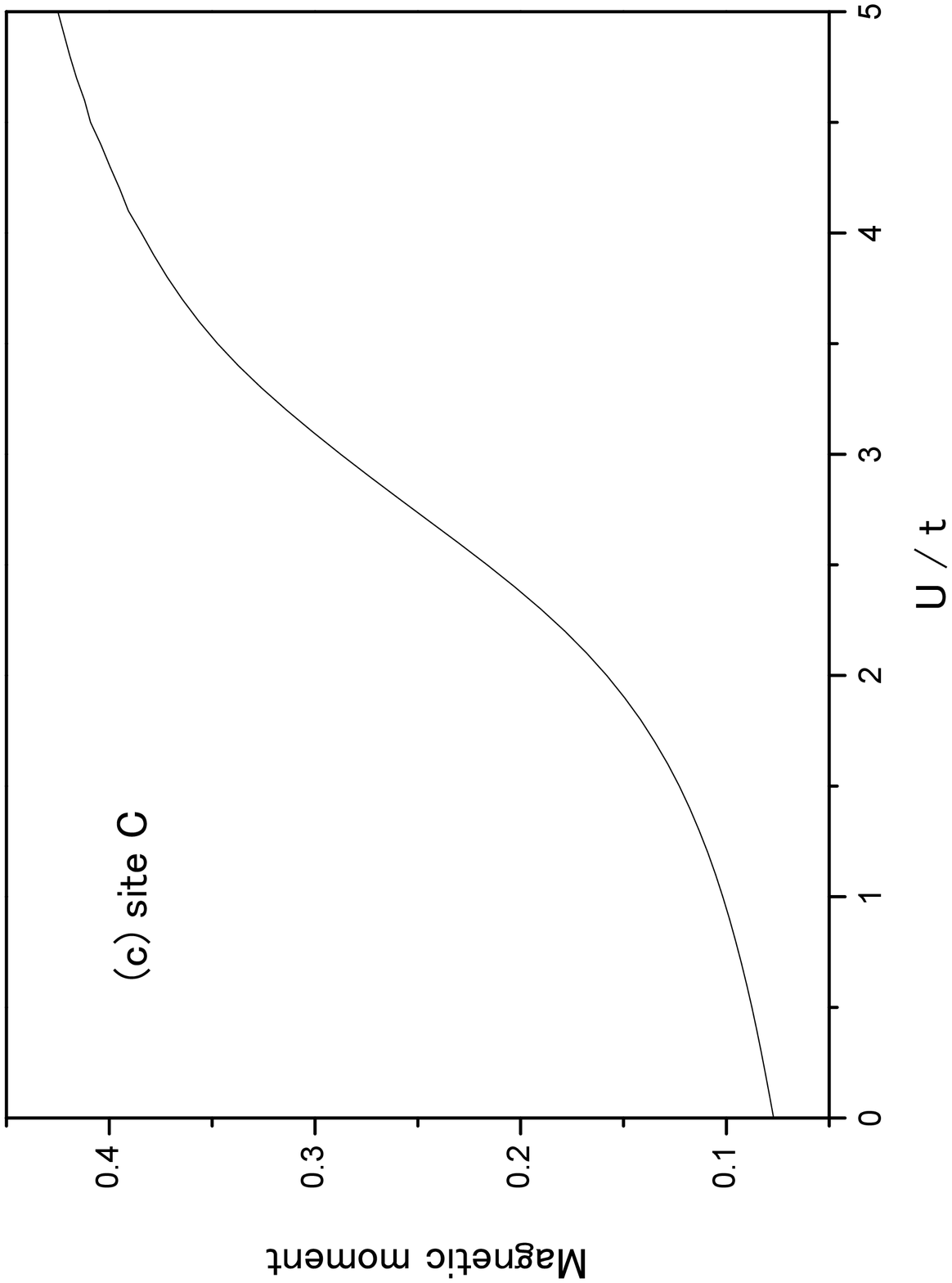}
}
\centerline{
\includegraphics[height=5cm]{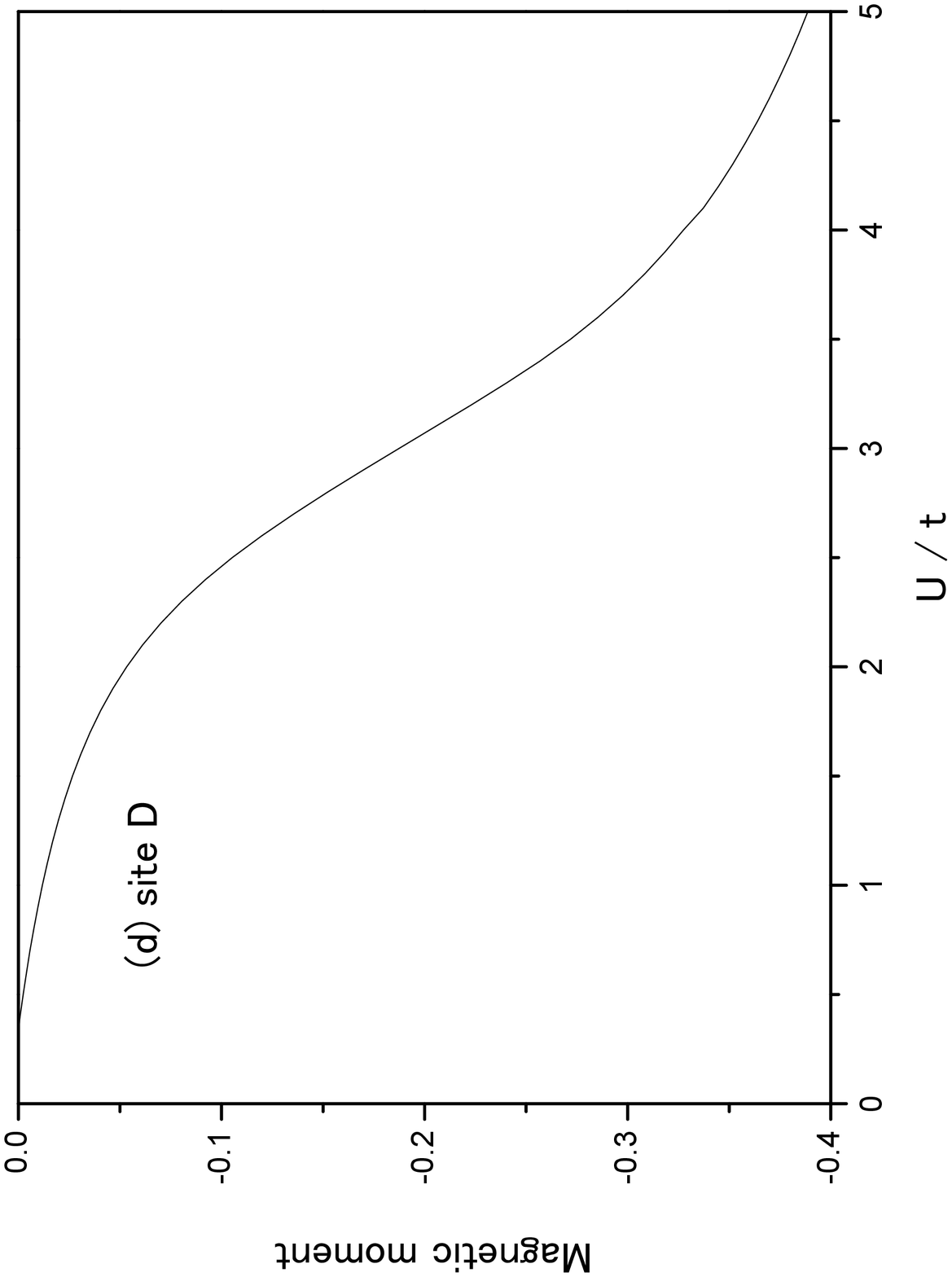}
}
\caption{
The magnetic moment at the site A (a), site B (b),
site C (c), and site D (d), as a function of the Coulomb interaction 
strength $U$ of a part of C$_{60}$.
}
\label{fig:Fig4}
\end{figure}

\end{document}